\documentclass[conference]{IEEEtran}
\IEEEoverridecommandlockouts
\usepackage{cite}
\usepackage{amsmath,amssymb,amsfonts}
\usepackage{algorithmic}
\usepackage{graphicx}
\usepackage{textcomp}
\usepackage{booktabs}
\usepackage{xcolor}
\usepackage{url}
\usepackage{longtable}
\usepackage[hidelinks, hyperfootnotes=true]{hyperref}
\def\BibTeX{{\rm B\kern-.05em{\sc i\kern-.025em b}\kern-.08em
    T\kern-.1667em\lower.7ex\hbox{E}\kern-.125emX}}
\def\best #1 {{\underline{#1}}}     

\begin{document}

\title{Comparative Analysis of Zero-Shot Capability of Time-Series Foundation Models in Short-Term Load Prediction
\thanks{* indicates equal contribution.}}

\author{\IEEEauthorblockN{\textbf{Nan Lin\textsuperscript{*1}, Dong Yun\textsuperscript{*2}, Weijie Xia\textsuperscript{*1}, Peter Palensky\textsuperscript{1}, Pedro P. Vergara\textsuperscript{1}}}

\IEEEauthorblockA{\textsuperscript{1} Intelligent Electrical Power Grids (IEPG), Delft University of Technology, Delft, The Netherlands}
\IEEEauthorblockA{\textsuperscript{2} Department of Electrical and Photonics Engineering, Technical University of Denmark, Kgs., Lyngby, Denmark}
\IEEEauthorblockA{Email: \{W.Xia, N.Lin, P.P.VergaraBarrios, P.Palensky\}@tudelft.nl, s232293@dtu.dk}
}


\maketitle

\begin{abstract}
Short-term load prediction (STLP) is critical for modern power distribution system operations, particularly as demand and generation uncertainties grow with the integration of low-carbon technologies, such as electric vehicles and photovoltaics. In this study, we evaluate the zero-shot prediction capabilities of five Time-Series Foundation Models (TSFMs)—a new approach for STLP where models perform predictions without task-specific training—against two classical models, Gaussian Process (GP) and Support Vector Regression (SVR), which are trained on task-specific datasets. Our findings indicate that even without training, TSFMs like Chronos, TimesFM, and TimeGPT can surpass the performance of GP and SVR. This finding highlights the potential of TSFMs in STLP.
\end{abstract}


\section{Introduction}
Short-term load prediction (STLP) is essential for the efficient operation of power distribution systems \cite{zhang2021review}. As modern power systems grow increasingly sophisticated, STLP serves as a fundamental tool for capturing dynamic fluctuations in load demand \cite{gong2019research}. Consequently, it has significant applications such as demand response \cite{wen2020load} and control optimization \cite{trudnowski2001real}. 




Before the prevalence of Deep Learning (DL), the STLP methods primarily relied on statistical models.  For instance, in \cite{huang2003short}, an Auto-Regressive Moving Average (ARMA) model is introduced for STLP that incorporates considerations for non-Gaussian processes. The method utilizes bispectral analysis to assess the Gaussianity of load data. Results indicate that this approach achieves much lower error than a standard ARMA model. In \cite{wang2021short, zhang2017short}, the Decomposition technologies are used to decompose load data first, and then models like XGBoost and Support Vector Machine (SVM) are applied to model the trend and sub-series. These hybrid ways show better performance than a single model. Given the significant impact of weather factors on load, the study in \cite{barman2018regional} proposes an SVM model enhanced with a Grasshopper Optimization algorithm to address STLP under specific weather conditions. This method performs better than other hybrid approaches, such as GA-SVM. The study in \cite{divina2018stacking} proposes an ensemble learning approach that uses models like regression trees and Random Forests as base learners. Like hybrid models, ensemble models consistently perform better than standalone models. The study in \cite{jiang2016short} introduces a Support Vector Regression (SVR) model with a customized grid traverse algorithm for high-resolution (1-second) STLP. This approach clearly outperforms traditional models in high-resolution STLP, such as GA-SVM and ARIMA.  In \cite{cao2021robust, zhao2023gaussian}, Gaussian Process (GP) models are applied for load prediction tasks. The integration of GP offers the advantage of reduced data requirements compared to standalone DL models.

However, with the growing availability of load data and DL advancements, DL-based STLP approaches have demonstrated superior performance compared to traditional statistical models in recent years. For instance, the study in \cite{lin2021spatial} introduces a Graph Neural Network (NN) for STLP. Unlike models that consider only temporal information, the Graph NN approach incorporates spatial information, recognizing that the load patterns of households within a region exhibit similarities. The study in \cite{sadaei2019short} proposes transforming load and weather data into an image format which allows Convolutional NN to be applied as forecasters. In \cite{fazlipour2022deep}, the author proposes a Long Short-Term Memory (LSTM) model with an Attention Mechanism, effectively capturing complex data correlations. This approach demonstrates better performance compared to a classical LSTM model. Despite the superior performance, one disadvantage of the above-mentioned models is they can only provide point prediction. The application of deep generative models in STLP tackles this issue. In \cite{wang2020modeling}, conditional Wasserstein GANs (cWGANs) are utilized for probabilistic load prediction, conditioned on weather data and historical load. This method outperforms traditional approaches, such as quantile regression. Additionally, studies in \cite{xia2024flow, lin2023residential} apply flow-based models for prediction, demonstrating that these models perform better than GANs and VAEs in probabilistic STLP. 

However, both statistical and DL-based models require training on specific datasets, which presents two challenges: (1) Despite increased data availability from metering devices, data for individual devices (e.g., households or substations) often remains limited, leading to potential data shortages during training \cite{song2023comprehensive}; (2) Training, particularly for DL models across various datasets, can be highly time-consuming. Time-Series Foundation Models (TSFM), characterized by the capability of zero-shot prediction, provide a potential solution to these challenges. The current prevalent TSFM models include Chronos \cite{ansari2024chronos}, Moment \cite{goswami2024moment}, Lag-llama \cite{rasul2024lag}, TimesFM \cite{das2023decoder},  and TimeGPT \cite{garza2023TimeGPT}, a detailed introduction of these TSFM is provided in Sec.~\ref{TSFM_IN}.  Most of these models are based on Transformer architectures and are pre-trained on large, diverse time series datasets from multiple domains for various prediction tasks. As such, they are designed to function as "general forecasters" similar to the Large Language Model (LLM). This allows them to be used without additional model training—particularly beneficial in data-scarce or zero-data scenarios. 


In this paper, we assess the zero-shot (meaning no training process for the TSFMs) prediction performance of state-of-the-art TSFM for STLP, benchmarking them against the widely adopted classical models, SVR and GP. This paper aims to explore a potential new research direction for the STLP task. To our knowledge, this study represents the first exploration of TSFMs in the context of load prediction. The code and data of this research are accessible in the project repository\footnote{\noindent Code and data are available in the following repositories: \\
1) \href{https://github.com/sentient-codebot/TSFM-RLP-Forecast/tree/main}{Personal Repository} \\
2) \href{https://github.com/sentient-codebot/TSFM-RLP-Forecast/tree/main}{TU Delft Project Repository (need to be given)}
}.

\section{Times-series Foundation Models}\label{TSFM_IN}
Since TSFMs share similar design principles, this section primarily focuses on introducing and comparing the differences between Chronos and Lag-llama to provide a comprehensive understanding of TSFM architectures.  For detailed information on model design, readers are encouraged to read the original papers \cite{ansari2024chronos,goswami2024moment,rasul2024lag,das2023decoder, garza2023TimeGPT}.

\begin{figure}[t]
    \centering
    \includegraphics[width=0.7\linewidth]{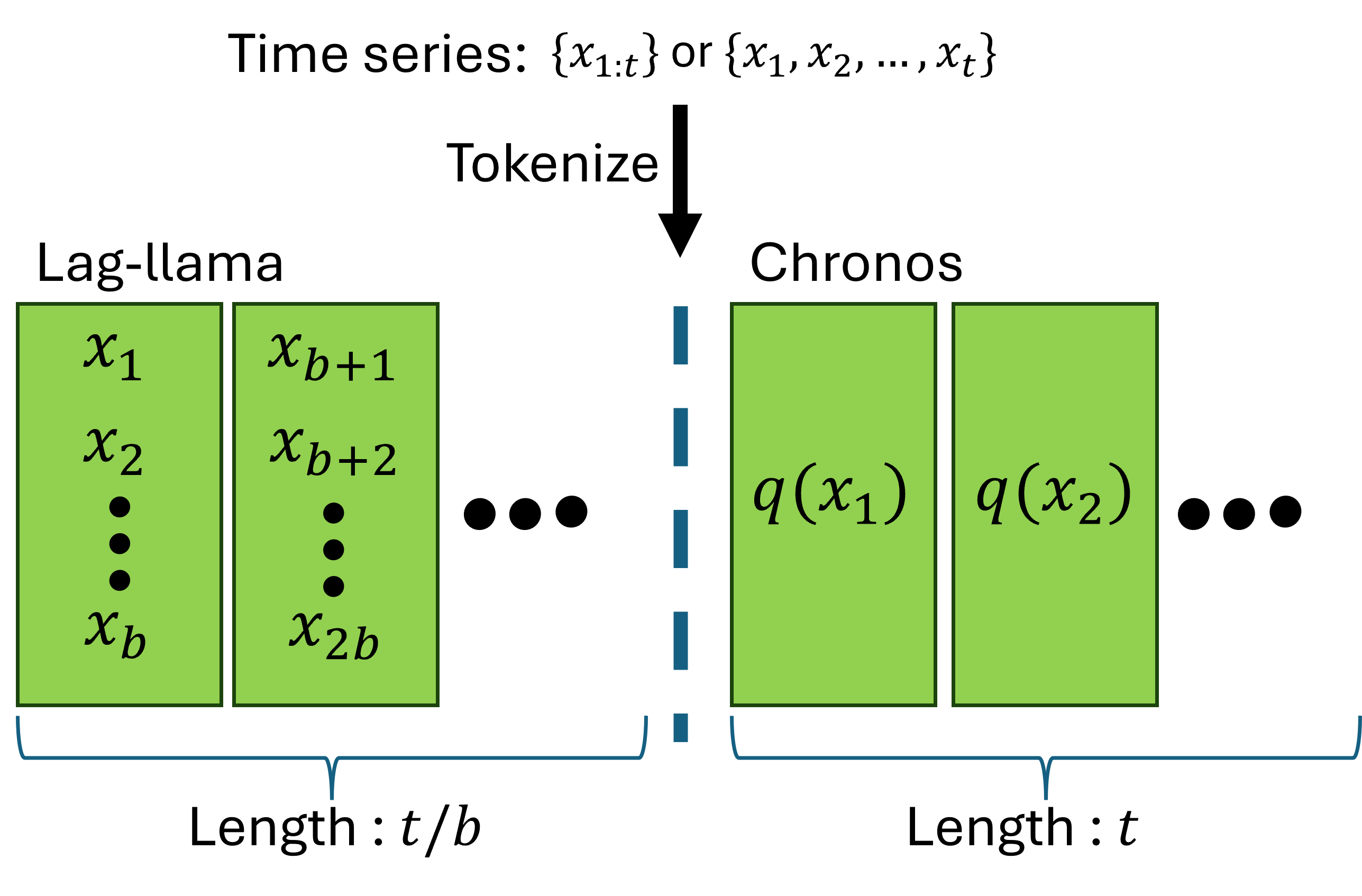}
    \caption{Two tokenization approaches for time series. $t$ is the length of the time series, and $b$ is the length of the token. The Lag-llama's approach is to segment time series equally to create a sequence of tokens (assume $t$ is divisible by $b$). The Chronos's approach is to map the real values to a token.}
    \label{tokenization}
\end{figure}

The primary difference between time series and language series lies in their data types: while a language series consists of discrete tokens representing specific words, a time series comprises a sequence of real values. The time series tokenization approach is commonly adopted in TSFM to adapt the Transformer architecture for time series and bridge the format gap between these data types.  For instance, in Lag-llama \cite{rasul2024lag}, the time series is segmented equally to create a sequence of tokens. In Chronos \cite{ansari2024chronos}, tokenization is achieved through a tokenization function $q(\cdot)$ that maps real value $x_i$ into a token $\mathbf{z_i}=q(x_i)$, Fig.~\ref{tokenization} illustrates these two tokenization approaches.

\begin{figure}[t]
    \centering
    \includegraphics[width=\linewidth]{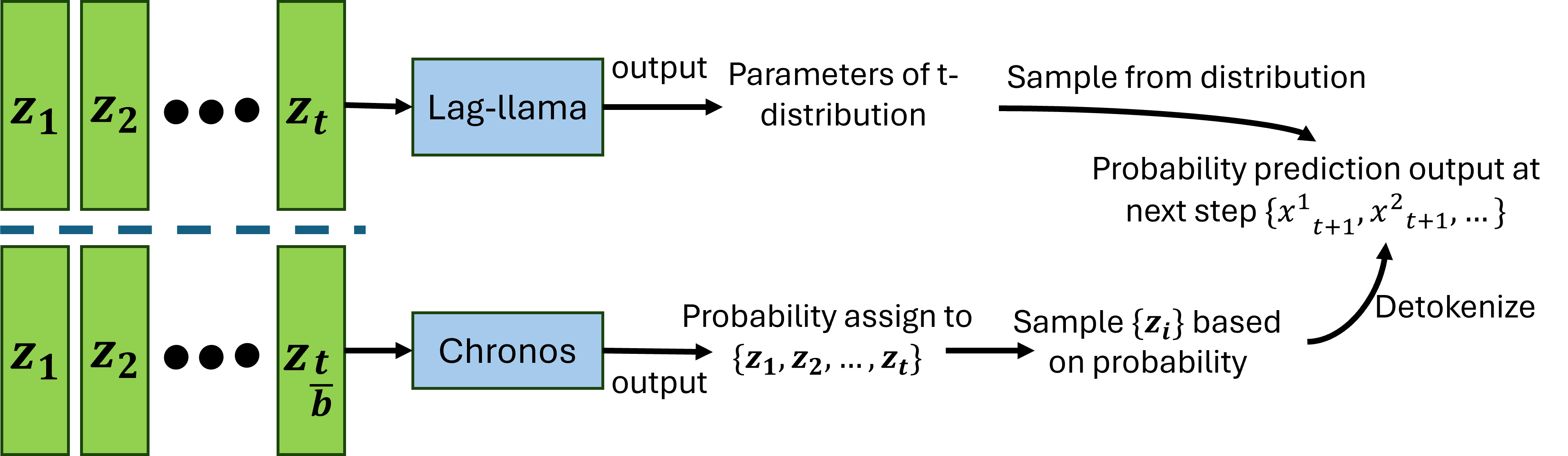}
    \caption{Two output mechanisms. $\mathbf{z_i}$ is the token which is obtained by tokenization. $\mathbf{z_i}=\{x_{i\times b+1},x_{i\times b+2},...,x_{i\times b+b}\}$ in Lag-llama, and $\mathbf{z_i}=q(x_i)$ in Chronos.}
    \label{output}
    \vspace{-0.5cm}
\end{figure}

\begin{figure*}
    \centering
    \includegraphics[width=0.70\linewidth]{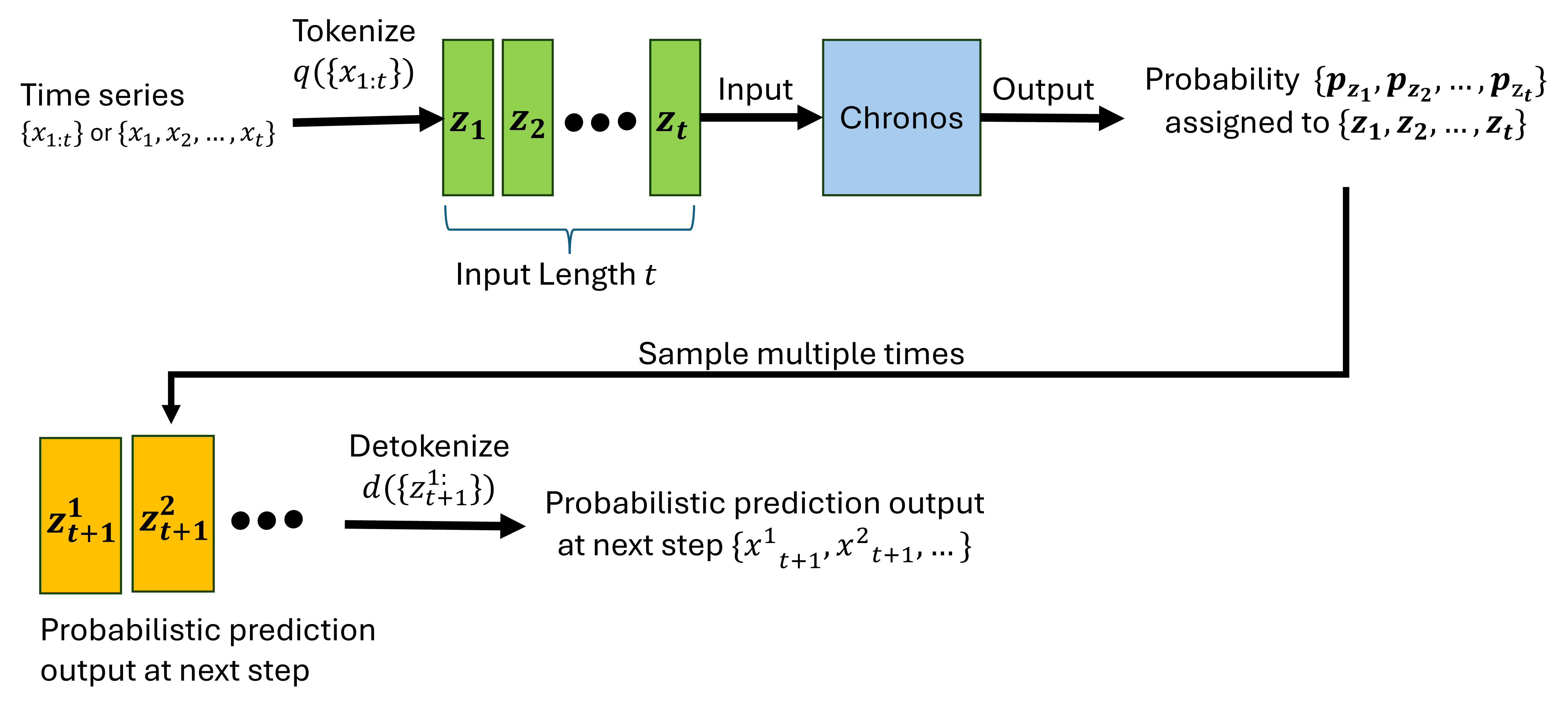}
    \vspace{-0.4cm}
    \caption{The prediction process of Chronos. $\{\mathbf{z}^{1:}_{t+1} \}$ is sampled from $\{ \mathbf{z}_{1:t}\}$ based on the probability $\{p_{\mathbf{z}_i}; i=1, 2,.., t \}$, where $\sum_i^t p_{\mathbf{z}_i} =1$.}
    \label{predprocess}
\end{figure*}
Once the time series is tokenized, a standard Transformer architecture can be applied to process the data. However, another challenge arises. Unlike LLMs, which output discrete tokens corresponding to specific words, the output of a TSFM is typically expected to be a sequence of continuous real values, reflecting the nature of the input time series data. To address this issue, Chronos adopts an approach similar to that of LLM, it outputs the probability distribution for each input token, and then samples input tokens based on these probabilities to obtain output tokens. The resulting tokens are subsequently detokenized using the detokenization function $d(\cdot)$ to generate predicted values $d(\mathbf{z})$. This method implicitly assumes that the time series remains stationary over time. In contrast, Lag-llama first predicts parameters of an exponential distribution rather than probabilities, then sampling values directly from this distribution as the prediction results. The output mechanism is shown in Fig.~\ref{output}. Moreover, due to the output of these two models is essentially a distribution (or probability), we can sample multiple times from the distribution to obtain the probabilistic prediction results. Fig~\ref{predprocess} illustrates a more detailed example of the TSFM model’s prediction process using Chronos, including both tokenization and detokenization steps.
\vspace{-0.3cm}




\begin{table}[t]
    \centering
    \caption{Datasets Used for the Model Comparison.}
    \begin{tabular}{ccccc}
    \hline
    \textbf{ID}  &  \textbf{Country} & \textbf{Resolution} & \textbf{Days} & \textbf{Date Range} \\
    \hline
    \multicolumn{5}{c}{\textit{Individual Load Data}}\\
    \hline
    NL-I-60  &  NL & 60-minutes & 15,600 & 01.2013-12.2013 \\
    GE-I-60  &  GE & 60-minutes & 1,200 & 12.2014-05.2019 \\
    GE-I-30  &  GE & 30-minutes & 1,200 & 12.2014-05.2019 \\
    GE-I-15  &  GE & 15-minutes & 1,200 & 12.2014-05.2019 \\
    UK-I-30  &  UK & 30-minutes & 7,200 & 01.2013-12.2013 \\
    UK-I-60  &  UK & 60-minutes & 7,200 & 01.2013-12.2013 \\
    \hline
    \multicolumn{5}{c}{\textit{Aggregated Load Data}}\\
    \hline
    NL-A-60  &  NL & 60-minutes & 600 & 01.2013-12.2013 \\
    GE-A-60  &  GE & 60-minutes & 200 & 12.2014-05.2019 \\
    GE-A-30  &  GE & 30-minutes  & 200 & 12.2014-05.2019 \\
    GE-A-15  &  GE & 15-minutes & 200 & 12.2014-05.2019 \\
    UK-A-30  &  UK & 30-minutes & 2,400 & 01.2013-12.2013 \\
    UK-A-60  &  UK & 60-minutes & 2,400 & 01.2013-12.2013 \\
    \hline
    \end{tabular}
    \label{dataset}
\begin{flushleft}
\fontsize{7}{8}\selectfont
\textbf{Days}: The total number of days for which
load data is available. 
\end{flushleft}
\vspace{-0.5cm}
\end{table}

\begin{table}[t]
    \centering
     \caption{Comparison of Model features.}
    \begin{tabular}{cccc}
        \hline
        \textbf{Model} & \textbf{Type} & \textbf{Probabilistic}  & \textbf{Scale}\\
        \hline
        Chronos-tiny  & TSFM  & Yes & 8M \\
        Chronos-small  & TSFM  & Yes & 46M \\
        Moment   & TSFM  & No &  37.9M\\
        TimeGPT   & TSFM & No & / \\
        TimeFM   & TSFM & No    & 200M \\
        Lag-llama   & TSFM  & YES & 2.45M \\
        SVR & Statistical Model & No & / \\
        GP & Statistical Model & YES & /\\
        \hline
    \end{tabular}
    \label{Models}

\begin{flushleft}
\fontsize{7}{8}\selectfont
\textbf{Probabilistic}: Whether the model supports probabilistic prediction. \\
\textbf{Scale}: The number of parameters of the model, "M" means millions. The scale of TimeGPT is not publicly released.
\end{flushleft}
\vspace{-0.7cm}
\end{table}

\section{Experimental Setting}

\subsection{Dataset} The dataset used in this study includes individual and aggregated electricity consumption data at 60-minute, 30-minute, and 15-minute resolutions from the UK \cite{uk_data}, Germany \cite{ge_data}, and the Netherlands \cite{liander}. Table~\ref{dataset} summarizes the characteristics of all datasets used in this research. The "Days" column in Table~\ref{dataset} indicates the total number of days for which load data is available.

\subsection{Experimental Design} Five TSFM models—Chronos, Moment, Lag-llama, TimesFM, and TimeGPT—are compared with SVR and GP as classical approaches for STLP. Table~\ref{Models} summarizes the key features of all models used. For Chronos, multiple pre-trained model versions are available \cite{ansari2024chronos}, and we selected the two versions shown in Table~\ref{Models}, which differ only in the number of model parameters. The electricity consumption data is divided into training and test sets. The TSFMs are supplied with three days of historical load data in the test sets, resulting in $72 \times 1$, $72 \times 2$, and $72 \times 4$ data points for 60-minute, 30-minute, and 15-minute resolutions, respectively. These models then forecast the load for the following day, producing $24 \times 1$, $24 \times 2$, and $24 \times 4$ data points. For the SVR and GP models, training is conducted on the training set (60\%), with testing on the test sets (40\%), allowing for direct comparison with the TSFMs.

\subsection{Evaluation} For evaluation metrics, we calculate the Mean Absolute Error (MAE) and Root Mean Squared Error (RMSE) between the predicted average values (or the single predicted value for models that do not support probabilistic predictions) and the actual values. These metrics are expressed below

\begin{equation}
\text{MAE} = \frac{1}{T} \sum_{t=1}^{T} |x_t - \hat{x}_t|, 
\end{equation}

\begin{equation}
\text{RMSE} = \sqrt{\frac{1}{T} \sum_{t=1}^{t} (x_t - \hat{x}_t)^2},
\end{equation}
where $x_t$ and $\hat{x}_t$ are the real load value and predicted (average) load value at time index $t$. $T$ is the prediction length.  For models that support probabilistic predictions, we additionally compute the quantile losses at the 10\%, 50\%, and 90\% levels. The metric for computing the quantile loss is expressed as 
\begin{equation}
\text{QL}_{\gamma} = \frac{1}{T} \sum_{i=1}^{T} \left( \mathbb{I}_{\hat{x}_t \geq x_t} (1 - \gamma) |z_t - \hat{x}_t| + \mathbb{I}_{\hat{x}_t < x_t} \gamma |x_t - \hat{x}_t| \right),
\end{equation}
where $\text{QL}_{\gamma}$ is the quantile loss at quantile level $\gamma \in \{ 0.1,0.5,  0.9\}$, $\mathbb{I}(\cdot)$ is Indicator function.

\begin{figure*}[htp]
    \centering
    \begin{minipage}{2\columnwidth}
        \centering
        \includegraphics[width=\columnwidth]{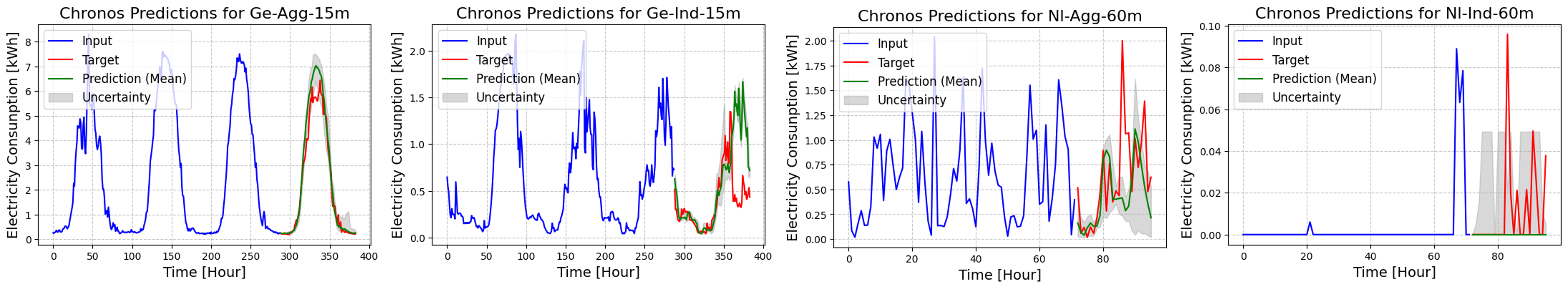}
        \text{\small (a) Prediction examples of Chronos-small}
        \vspace{6pt}
        \label{chronos}
    \end{minipage}
    \\

    \begin{minipage}{2\columnwidth}
        \centering
        \includegraphics[width=\columnwidth]{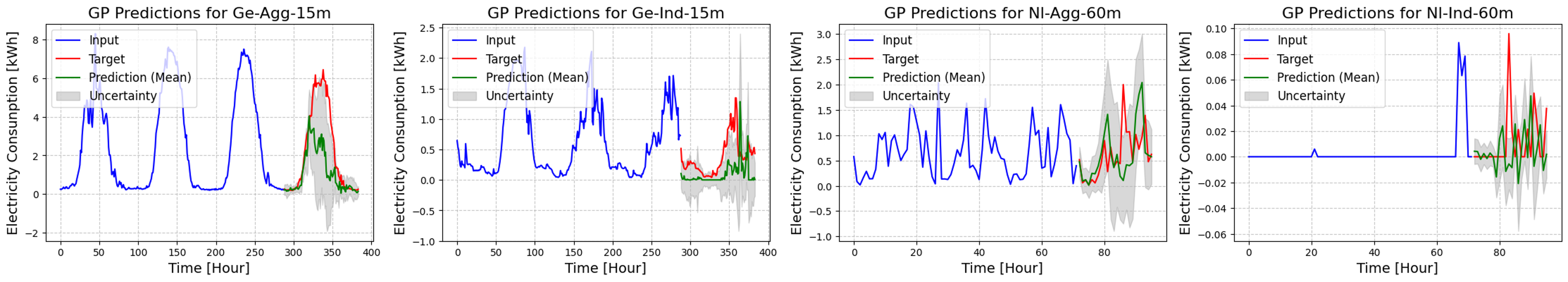}
        \text{\small (b) Prediction examples of GP}
        \vspace{6pt}
        \label{gp}
    \end{minipage}
    \\

    \begin{minipage}{2\columnwidth}
        \centering
        \includegraphics[width=\columnwidth]{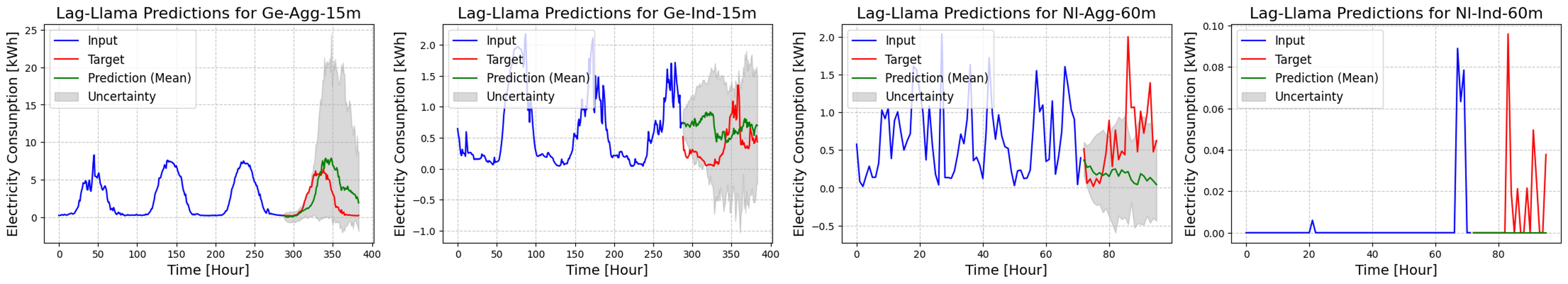}
        \text{\small (c) Prediction examples of Lag-llama}
        \label{lag}
    \end{minipage}
    \\


    \begin{minipage}{2\columnwidth}
        \centering
        \includegraphics[width=\columnwidth]{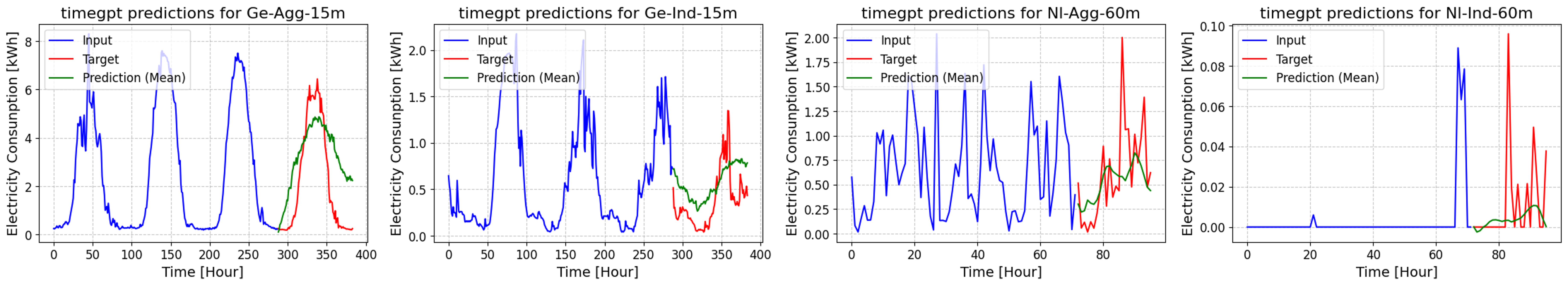}
        \text{\small (d) Prediction examples of TimeGPT}
        \label{lag}
    \end{minipage}
    \\
    
    \caption{Prediction examples of GP, TimeGPT, Lag-llama, and Chronos.}
    \label{pro_pred}
\end{figure*}

\begin{table}[t]
\centering
\caption{Experiential Results of Aggregated Load Data}
\begin{tabular}{cccccc}
\hline
\textbf{Model} & \textbf{Q 10\%} & \textbf{Q 50\%} & \textbf{Q 90\%} &\textbf{ MAE} & \textbf{RMSE} \\
\hline
\multicolumn{6}{c}{\textit{Experiment ID : GE-A-15}} \\
\hline
Moment & / & / & / & 1.3186 & 1.5784 \\
GP & 0.1969 & 0.4796 & 0.2676 & 0.9592 & 2.2039 \\
SVR & / & / & / & 0.8406 & 1.6161 \\
Chronos-small & 0.1014 & 0.2258 & 0.1540 & 0.4517 & 0.7040 \\
Chronos-tiny & \best 0.0968 & \best 0.2171 & \best 0.1503 & \best 0.4343 & \best 0.6863 \\
Lag-llama & 0.2968 & 0.5885 & 0.5249 & 1.1771 & 1.6455 \\
TimeGPT & / & / & / & 1.0419 & 1.2109 \\
TimesFM & / & / & / & 1.1747 & 1.6302 \\
\hline
\multicolumn{6}{c}{\textit{Experiment ID : GE-A-30}} \\
\hline
Moment & / & / & / & 2.6974 & 3.1888 \\
GP & \best 0.3876 & 0.9202 & 0.5104 & 1.8404 & 2.8721 \\
SVR & / & / & / & 1.6922 & 2.5012 \\
Chronos-small & 0.4834 & \best 0.6457 & 0.4023 & \best 1.2914 & 1.9006 \\
Chronos-tiny & 0.4719 & 0.6524 & \best 0.3895 & 1.3048 & 1.9307 \\
Lag-llama & 0.4999 & 1.0307 & 0.7717 & 2.0614 & 2.8865 \\
TimeGPT & / & / & / & 2.4966 & 2.9227 \\
TimesFM & / & / & / & 1.3126 & \best 1.8802 \\
\hline
\multicolumn{6}{c}{\textit{Experiment ID : GE-A-60}} \\
\hline
Moment & / & / & / & 5.3237 & 6.3017 \\
GP & 0.7755 & 1.8548 & 1.0545 & 3.7095 & 5.7497 \\
SVR & / & / & / & 3.4079 & 5.4665 \\
Chronos-small & 0.4835 & 0.8018 & 0.3894 & 1.6037 & 2.4774 \\
Chronos-tiny & \best 0.4616 & \best 0.7859 & \best 0.3859 & \best 1.5718 & \best 2.4190 \\
Lag-llama & 0.6847 & 1.9988 & 1.6902 & 3.9977 & 5.4701 \\
TimeGPT & / & / & / & 2.0881 & 2.9514 \\
TimesFM & / & / & / & 2.1064 & 3.1958 \\
\hline
\multicolumn{6}{c}{\textit{Experiment ID : UK-A-30}} \\
\hline
Moment & / & / & / & 0.2262 & 0.2947 \\
GP & 0.2602 & 0.1176 & 0.2971 & 0.2353 & 0.4246 \\
SVR & / & / & / & 0.2591 & 0.3133 \\
Chronos-small & 0.0217 & \best 0.0691 & 0.0868 & \best 0.1383 & 0.2392 \\
Chronos-tiny & \best 0.0213 & 0.0692 & \best 0.0862 & \best 0.1383 & 0.2382 \\
Lag-llama & 0.0452 & 0.1192 & 0.1351 & 0.2384 & 0.3276 \\
TimeGPT & / & / & / & 0.1456 & \best 0.2066 \\
TimesFM & / & / & / & 0.1560 & 0.2317 \\
\hline
\multicolumn{6}{c}{\textit{Experiment ID : UK-A-60}} \\
\hline
Moment & / & / & / & 0.2299 & 0.3002 \\
GP & 0.0448 & 0.0852 & \best 0.0536 & 0.1703 & 0.2386 \\
SVR & / & / & / & 0.1801 & 0.2190 \\
Chronos-small & 0.0222 & \best 0.0717 & 0.0826 & \best 0.1433 & 0.2324 \\
Chronos-tiny & \best 0.0221 & 0.0720 & 0.0828 & 0.1441 & 0.2407 \\
Lag-llama & 0.0390 & 0.1133 & 0.1231 & 0.2266 & 0.3163 \\
TimeGPT & / & / & / & 0.1504 & 0.2202 \\
TimesFM & / & / & / & 0.1449 & \best 0.2188 \\
\hline
\multicolumn{6}{c}{\textit{Experiment ID : NL-A-60}} \\
\hline
Moment & / & / & / & 2.7385 & 3.1118 \\
GP & 0.3064 & 0.6320 & \best 0.3115 & 1.2639 & 1.5064 \\
SVR & / & / & / & 1.3312 & 1.6444 \\
Chronos-small & 0.2578 & 0.5330 & 0.3974 & 1.0660 & 1.3988 \\
Chronos-tiny & \best 0.2471 & \best 0.5290 & 0.3980 & \best 1.0580 & 1.3936 \\
Lag-llama & 0.8456 & 1.8532 & 1.7357 & 3.7064 & 4.1879 \\
TimeGPT & / & / & / & 1.0879 & \best 1.3697 \\
TimesFM & / & / & / & 1.1175 & 1.4481 \\
\hline

\hline
\label{exp_results_agg}
\end{tabular}
\begin{flushleft}
\fontsize{7}{8}\selectfont
\textbf{Q $h$\%}: Meaning the quantile loss at quantile level $h$.
\end{flushleft}
\vspace{-0.7cm}
\end{table}

\begin{table}[t]
\centering
\caption{Experiential Results of Individual Load Data}
\begin{tabular}{cccccc}
\hline
\textbf{Model} & \textbf{Q 10\%} & \textbf{Q 50\%} & \textbf{Q 90\%} &\textbf{MAE} & \textbf{RMSE} \\
\hline
\multicolumn{6}{c}{\textit{Experiment ID : GE-I-15}} \\
\hline
Moment & / & / & / & 0.2761 & 0.3513 \\
GP & 0.0475 & 0.1216 & 0.0831 & 0.2432 & 0.3498 \\
SVR & / & / & / & 0.2415 & 0.3463 \\
Chronos-small & 0.0319 & 0.0716 & 0.0687 & 0.1433 & \best 0.2202 \\
Chronos-tiny & \best 0.0289 & \best 0.0700 & \best 0.0679 & \best 0.1399 & 0.2231 \\
Lag-llama & 0.0526 & 0.1425 & 0.1290 & 0.2849 & 0.3730 \\
TimeGPT & / & / & / & 0.1845 & 0.2313 \\
TimesFM & / & / & / & 0.1779 & 0.2622 \\
\hline
\multicolumn{6}{c}{\textit{Experiment ID : GE-I-30}} \\
\hline
Moment & / & / & / & 0.5183 & 0.6636 \\
GP & 0.0905 & 0.2574 & 0.2173 & 0.5148 & 0.7506 \\
SVR & / & / & / & 0.4267 & 0.6020 \\
Chronos-small & 0.0550 & \best 0.1410 & \best 0.1366 & \best 0.2821 & 0.4593 \\
Chronos-tiny & \best 0.0497 & 0.1422 & 0.1375 & 0.2845 & 0.4636 \\
Lag-llama & 0.0818 & 0.2362 & 0.2318 & 0.4723 & 0.6644 \\
TimeGPT & / & / & / & 0.4108 & 0.5106 \\
TimesFM & / & / & / & 0.2947 & \best 0.4177 \\
\hline
\multicolumn{6}{c}{\textit{Experiment ID : GE-I-60}} \\
\hline
Moment & / & / & / & 1.0095 & 1.2799 \\
GP & 0.1845 & 0.5346 & 0.4401 & 1.0692 & 1.2763 \\
SVR & / & / & / & 0.8028 & 1.1808 \\
Chronos-small & 0.1122 & 0.2232 & \best 0.1636 & 0.4464 & 0.7146 \\
Chronos-tiny & \best 0.0945 & \best 0.2189 & 0.1656 & \best 0.4378 & 0.7034 \\
Lag-llama & 0.1504 & 0.5067 & 0.5046 & 1.0135 & 1.3523 \\
TimeGPT & / & / & / & 0.4807 & 0.6693 \\
TimesFM & / & / & / & 0.4428 & \best 0.6633 \\
\hline
\multicolumn{6}{c}{\textit{Experiment ID : UK-I-30}} \\
\hline
Moment & / & / & / & 0.1847 & 0.2428 \\
GP & 0.1774 & 0.1644 & 0.2123 & 0.3287 & 1.0757 \\
SVR & / & / & / & 0.2262 & 0.2989 \\
Chronos-small & 0.0196 & \best 0.0568 & 0.0662 & \best 0.1136 & \best 0.1748 \\
Chronos-tiny & \best 0.0190 & 0.0569 & \best 0.0658 & 0.1138 & 0.1935 \\
Lag-llama & 0.0362 & 0.0899 & 0.0956 & 0.1798 & 0.2501 \\
TimeGPT & / & / & / & 0.1273 & 0.1799 \\
TimesFM & / & / & / & 0.1194 & 0.1795 \\
\hline
\multicolumn{6}{c}{\textit{Experiment ID : UK-I-60}} \\
\hline
Moment & / & / & / & 0.2685 & 0.3419 \\
GP & 0.0625 & 0.1322 & 0.1076 & 0.2645 & 0.3680 \\
SVR & / & / & / & 0.2634 & 0.3368 \\
Chronos-small & \best 0.0240 & 0.0746 & 0.0860 & 0.1492 & \best 0.2257 \\
Chronos-tiny & \best 0.0240 & \best 0.0745 & \best 0.0858 & \best 0.1490 & 0.2469 \\
Lag-llama & 0.0496 & 0.1320 & 0.1422 & 0.2640 & 0.3547 \\
TimeGPT & / & / & / & 0.1588 & 0.2313 \\
TimesFM & / & / & / & 0.1505 & 0.2263 \\
\hline
\multicolumn{6}{c}{\textit{Experiment ID : NL-I-60}} \\
\hline
Moment & / & / & / & 0.2641 & 0.3237 \\
GP & 0.0456 & 0.1022 & \best 0.0662 & 0.2044 & 0.2869 \\
SVR & / & / & / & 0.2239 & 0.2675 \\
Chronos-small & \best 0.0219 & \best 0.0710 & 0.0726 & 0.1420 & 0.2305 \\
Chronos-tiny & \best 0.0219 & \best 0.0710 & 0.0725 & 0.1421 & 0.2306 \\
Lag-llama & 0.0421 & 0.1173 & 0.1392 & 0.2345 & 0.3227 \\
TimeGPT & / & / & / & 0.1416 & \best 0.1950 \\
TimesFM & / & / & / & \best 0.1397 & 0.2001 \\
\hline

\hline
\label{exp_results_ind}
\end{tabular}
\begin{flushleft}
\fontsize{7}{8}\selectfont
\textbf{Q $h$\%}: Meaning the quantile loss at quantile level $h$.
\end{flushleft}
\vspace{-0.7cm}

\end{table}

\section{Results and Discussion}\label{experiment}
Tables~\ref{exp_results_agg} and~\ref{exp_results_ind} present the experimental results for aggregated and individual datasets, respectively. Chronos consistently achieves superior performance in probabilistic prediction across both dataset types, attaining the lowest quantile losses at the 10\%, 50\%, and 90\% levels. Overall, Chronos-small exhibits a slight performance advantage over Chronos-tiny, likely due to its larger model size as shown in Table~\ref{Models}. Notably, despite training, GP does not outperform Chronos in any experiment except NL-I-60. Even compared with Lag-llama, a relatively weaker TSFM, GP fails to demonstrate clear advantages. For instance, Lag-llama outperforms GP in the NL-A-60, UK-A-60, and UK-I-30 examples.

Fig.~\ref{pro_pred} (a-d) provides prediction examples of prediction results from Chronos, GP, Lag-llama, and TimeGPT. As illustrated in Fig.~\ref{pro_pred} (b), GP frequently predicts negative values, which theoretically do not exist in the dataset (as load values are always positive). This issue, which contributes to higher quantile losses, arises from GP’s reliance on the assumption that time series follow a multivariate Gaussian distribution—a simplification inadequate for modeling load pattern volatility, especially in individual datasets. In contrast, Chronos, as shown in Fig.~\ref{pro_pred} (a), avoids these inaccuracies. Additionally, Chronos more accurately captures peak load values in the Ge-A-15 and GE-I-15 examples than GP. However, we observed that Chronos, along with nearly all models, fails to accurately predict the load for the highly volatile IN-I-60 example shown in Fig. \ref{pro_pred}.

Lag-llama, as another TSFM model, performs worse than Chronos in terms of probabilistic prediction, as indicated in Tables~\ref{exp_results_agg} and~\ref{exp_results_ind}. Fig.~\ref{pro_pred} (c) shows that, similar to GP, Lag-llama generates negative values with even larger variation (reflected by broad uncertainty intervals) and diverges notably in the NL-A-60 example. Lag-llama's weaker performance can be attributed to (1) its smaller scale, evident in the NL-A-60 example in Fig.~\ref{pro_pred} (c), where Lag-llama fails to capture the load data pattern, and (2) its reliance on a t-distribution assumption, which inevitably generates negative values similar to GP, as seen in the GE-A-15 and GE-I-15 examples in Fig.~\ref{pro_pred} (c). This limitation stems from the long-tailed nature of the t-distribution, which may be inadequate for modeling load data. In contrast, Chronos's good performance could first be attributed to its tokenization methods. In Chronos's prediction process, as shown in Fig.~\ref{predprocess} Chronos's prediction is essentially randomly sampled tokens from the historical tokens, this will ensure that the predicted load values are non-negative (as there are no negative historical load values). However, this tokenization strategy might affect Chronos's ability to predict unseen peak values.

In terms of point prediction, GP and SVR—classical models that require training—do not show advantages over TSFMs, consistently yielding relatively higher MAE and RMSE values. Chronos continues to demonstrate the strongest performance in point prediction. Additionally, TimesFM and TimesGPT exhibit competitive results, frequently achieving the lowest error rates, as seen in several scenarios. In contrast, Moment and Lag-llama generally show weaker performance in point prediction. Moment employs tokenization methods similar to those used in Lag-llama \cite{goswami2024moment}. Given that Moment has a large scale among the TSFMs discussed, it is expected to deliver strong performance. However, its comparatively weaker results may be attributed to the fact that, unlike other TSFMs specifically designed for predictive tasks, Moment is intended as a foundational model capable of supporting diverse tasks, such as classification and detection.


In summary, Chronos demonstrates competitive performance in probabilistic prediction compared to classical models like GP, even without training. By contrast, Lag-llama does not show clear advantages over GP. For point prediction, Chronos maintains superior accuracy, with TimeGPT also performing well. Moment appears to predict only average values across experiments, while SVR, GP, and Lag-llama generally show weaker performance than Chronos and TimeGPT.



\section{Conclusion}
In this paper, we examine the zero-shot performance of five TSFMs including Chronos, Moment, TimesFM, Lag-llama, and TimeGPT comparing them to two classical models, GP and SVR. The experimental results indicate that, under the same conditions, pre-trained TSFMs such as TimeGPT and Chronos can outperform classical models that are trained on the dataset. We observe, that TSFMs, especially Chronos show a strong performance in both probabilistic and point prediction.

We consider TSFMs to be a promising approach for future STLP tasks, as once trained, TSFMs require little to no additional data for specific training, effectively addressing data scarcity for data privacy or data collection issues without compromising much performance.

However, current TSFMs cannot incorporate external conditions, which is their primary limitation compared to other models. As shown in \cite{barman2018regional}, external conditions and information can effectively improve prediction accuracy. Therefore, a key direction for future research is to integrate external conditions, such as weather information, into the forecasting mechanism.

\section*{Acknowledgments}
This publication is part of the project ALIGN4Energy (with project number NWA.1389.20.251) of the research programme NWA ORC 2020 which is (partly) financed by the Dutch Research Council (NWO). This research utilized the Dutch National e-Infrastructure with support from the SURF Cooperative (grant number: EINF-5398).

\bibliographystyle{IEEEtran}
\bibliography{bib}

\end{document}